\documentclass{article}
\usepackage{spconf,amsmath,amssymb,graphicx}
\usepackage{xcolor}

\usepackage{booktabs}
\usepackage{graphicx}
\usepackage{makecell}
\usepackage{multirow}
\usepackage{tikz}
\usepackage[hidelinks]{hyperref}

\hypersetup{
    colorlinks=false,
    pdfborder={0 0 0}
}

\newcommand{\blackcircled}[1]{%
  \tikz[baseline=(char.base)]{
    \node[shape=circle,fill=black,text=white,inner sep=1pt] (char) {#1};
  }%
}
\newcommand{\sys}{SteerSpeech}
\title{SteerSpeech: Activation Steering for Emotion Control in generated speech}
\name{Afsara Benazir$^{1,2}$, Darius P\'etermann$^{2}$, Felix Xiaozhu Lin$^{1}$, Salar Rahili$^{2}$\thanks{This work was performed while A.~Benazir was an intern at Netflix.}}
\address{
$^{1}$University of Virginia\\
$^{2}$Netflix, Inc.
}
\begin{document}
\raggedbottom
\ninept
\maketitle

\begin{abstract}
Pretrained text-to-speech (TTS) models can generate expressive speech, but reliable inference-time emotion control remains challenging: prompts and reference audio offer coarse, inconsistent control, whereas specialized conditioning and model adaptation require costly training. 
We present \sys{}, a lightweight activation-steering framework that controls emotion by injecting steering vectors into hidden activations. For each target emotion we train a lightweight low-rank transform, using a multi-expert objective that
encourages monotonic emotion control while preserving speaker identity and linguistic content, constraining steering drift, and keeping the TTS backbone frozen.
To optimize through discrete speech tokens, we introduce a two-pass generation-and-replay pipeline using a straight-through estimator to backpropagate expert supervision through sampled tokens.
At inference, a target-emotion steering direction is optimized with its respective transform and injected into the base TTS model. Objective and subjective evaluations with Qwen3-TTS across seen, unseen, and accented speakers show stronger continuous emotion control with limited speaker and content degradation. SteerSpeech achieves 1.08$\times$–7.12$\times$ baseline target-emotion scores and for a representative emotion subjectively, it receives 78.1\%–96.8\% intensity preference and 1.43$\times$–1.46$\times$ speaker-identity preservation at high steering strengths.

\end{abstract}

\begin{figure*}[t]
    \centering
    \includegraphics[width=0.95\linewidth]{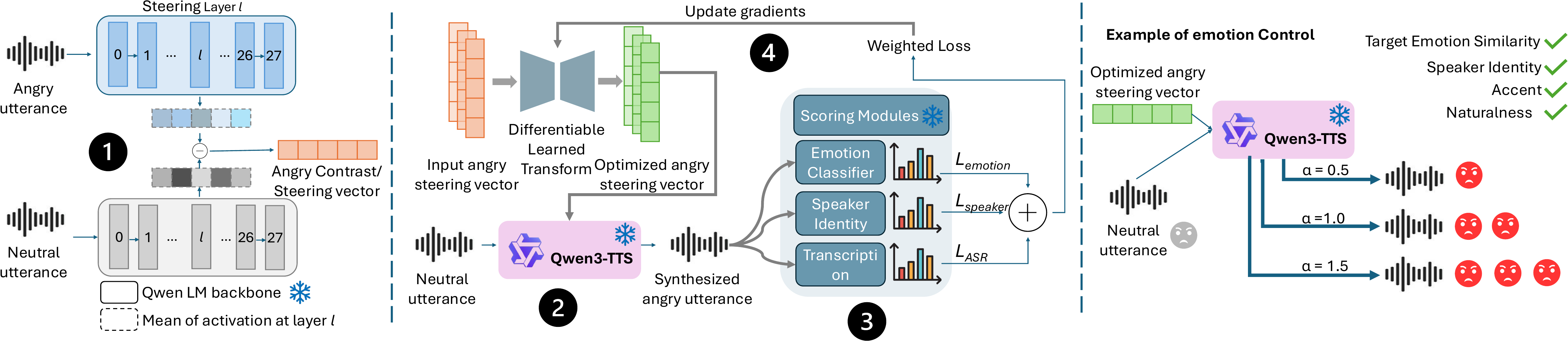}
    \caption{\sys{} overview.}
    \label{fig:o}
\end{figure*}

\section{Introduction}
\label{sec:intro}

Modern text-to-speech (TTS) systems generate highly natural speech, yet precise inference-time control of emotion remains challenging. Existing approaches condition synthesis on emotion labels, reference speech, natural-language descriptions, or prosodic controls~\cite{inoue2024hierarchical,jing2025paraevits,chen2025drawspeech}. Such controls can be coarse or inconsistent, while introducing a specialized encoder or adapting the TTS model for emotion controllability requires computationally expensive training and substantial labeled data. A more practical approach should enable downstream modification of attributes such as emotion, accent, and speaking style without retraining the underlying TTS model.

Activation steering \cite{rimsky2024steering} provides a lightweight alternative by modifying a pretrained model’s intermediate representations using a steering direction derived from the difference between mean of target and neutral activations. Prior work applies this approach to emotional speech synthesis:
EmoSteer-TTS \cite{xie2025emosteer} derives steering directions by identifying and modifying activation dimensions most relevant to the target emotion, whereas EmoShift \cite{zhou2026emoshift} transforms hidden representations into emotion-specific steering directions through learned mappings; EmoKnob \cite{chen2024emoknob} directly manipulates speaker embeddings.
However, the success of such intervention depends on how the steering direction is constructed, where it is applied, and how strongly it is injected. Steering directions may provide weak or unstable control, capture unrelated properties, 
alter speaker identity, linguistic content, or naturalness \cite{tan2024reliability} and vary with the steering strength, making continuous control difficult 
\cite{xie2025emosteer}.

Our \textbf{goal} is to enable continuous and fine-grained inference-time control of emotion in generated speech without retraining the TTS backbone or altering other attributes such as accent and speaking style.
Our key \textbf{insight} is that reliable emotion steering requires more than strengthening the target emotion: the steering direction should be jointly guided by emotion, speaker, and transcription experts to preserve identity and linguistic content. 

To that end, we \textbf{propose} \sys{}, a lightweight framework that learns a low-rank residual transform using multi-expert supervision enabled by a two-pass generation-and-replay pipeline that backpropagates through discrete speech tokens.
Given a coarse emotion contrast, i.e. the activation difference between target and neutral emotions,
the transform produces an optimized steering direction that strengthens the target emotion while preserving other speech attributes. 
We jointly optimize the transform across multiple steering strengths to encourage a monotonic increase in emotion intensity while limiting unpredictable behavior and off-target drift, all while keeping the TTS backbone frozen.
Our \textbf{contributions} are: 
\begin{itemize}
    \item We introduce \sys{}, a lightweight activation-steering framework for fine-grained, continuous inference-time control of emotion in generated speech without retraining the underlying TTS model.

    \item We learn a low-rank residual transform of emotion-steering directions using a multi-expert objective that explicitly promotes monotonic emotion control while preserving speaker identity, linguistic content, and speech quality.

    \item We demonstrate through objective and subjective evaluations that \sys{} provides fine-grained emotion controllability compared to baselines and generalizes across seen, unseen, and accented speakers.
\end{itemize}

\vspace{-3mm}
\section{Background}

\noindent\textbf{Emotion control in TTS.}
Existing systems guide emotional expression through conditioning signals at multiple levels. Categorical emotion labels provide simple utterance-level control but represent emotion using a small set of predefined classes~\cite{kang2023zetspeech}.
Reference speech transfers expressive characteristics from example speech~\cite{wang2018style}, while natural-language descriptions provide a more flexible interface for specifying emotional style~\cite{jing2025paraevits}. 
Finer control uses hierarchical emotion distributions~\cite{inoue2024hierarchical} or pitch and energy sketches for local prosodic trajectories~\cite{chen2025drawspeech}. However, these methods 
often require task-specific conditioning modules, joint TTS training, preference-based fine-tuning~\cite{gao2025emodpo}, or substantial backbone adaptation, increasing parameter and computational costs.

\noindent\textbf{Activation steering for controllable emotion synthesis.}
Activation steering~\cite{rimsky2024steering} offers a lightweight mechanism for controlling a pretrained model by modifying its hidden representations while keeping the model parameters frozen. Given a hidden state $h^{(l)}$ at layer $l$, a target-emotion direction is added with steering strength $\alpha$:
\begingroup
\setlength{\abovedisplayskip}{7pt}
\setlength{\belowdisplayskip}{7pt}
\setlength{\abovedisplayshortskip}{7pt}
\setlength{\belowdisplayshortskip}{7pt}
\begin{equation}
    v_e^{(l)}=\bar{h}_e^{(l)}-\bar{h}_{n}^{(l)}, \qquad
    \widetilde{h}^{(l)}=h^{(l)}+\alpha v_e^{(l)},
\end{equation}
\endgroup
where $\bar{h}_e^{(l)}$ and $\bar{h}_{n}^{(l)}$ denote the mean target-emotion and neutral activations, averaged over all decoding frames in respective utterances. Prior systems realize this intervention differently: 
EmoSteer-TTS~\cite{xie2025emosteer} uses neutral-to-emotion activation differences, while EmoShift~\cite{zhou2026emoshift} uses learned emotion-specific offsets. 
CoCoEmo~\cite{wang2026cocoemo} identifies effective steering sites and
composes mean-difference directions for single and mixed emotions, whereas HE-Vector~\cite{feng2026taskvector} constructs and combines
emotion and dialect vectors in model parameter space.  
In contrast, \sys{} uses explicit multi-expert supervision for monotonic emotion intensity and attribute preservation and learns a low-rank transform that optimizes a target steering direction at an internal layer.

\section{System Design}
\label{sec:sd}

\subsection{Overview}
\label{sec:overview}

We propose \sys{}, a lightweight, plug-and-play framework that learns to refine activation-steering directions for controllable emotion synthesis.

\noindent\textbf{Workflow}, is illustrated in Fig.~\ref{fig:o}. For each target emotion, we train a lightweight transform. Training consists of four steps: \blackcircled{1} extracting raw emotion contrasts from matched emotion and neutral utterance pairs, \blackcircled{2} transforming the contrast and injecting it into a frozen speech generator, \blackcircled{3} scoring the synthesized speech with frozen emotion, speaker, and transcription experts, and \blackcircled{4} updating only the learned transform, $T_\theta$ using their weighted objective function (Eq. \ref{eq:total_loss}). Except for $T_\theta$, all components of the underlying speech generation pipeline remain frozen (Sec. \ref{sec:training}).

\noindent\textbf{Inference.} Given a raw emotion contrast \(v_e\) i.e., mean activation difference between target-emotion and neutral utterances, \(T_\theta\) takes \(v_e\) as input and produces a refined steering direction. Steering with \(T_\theta(v_e)\) produces progressively stronger target emotion as the steering strength \(\alpha\) increases, while preserving speaker identity and linguistic content.

\noindent\textbf{System Components.}
\label{sec:components}
\sys{} consists of the following:

\noindent\textit{(1) Frozen speech generator.}
We use a pretrained autoregressive TTS model conditioned on target text and reference speech that predicts discrete speech tokens, which are subsequently decoded into waveform audio. 
We steer at an intermediate hidden layer following insights from \cite{wang2026cocoemo}. 

\noindent\textit{(2) Low-rank learned transform.}
The transform is the only trainable component: it normalizes a raw emotion contrast $v_e$ and applies a rank-$r$ normalized-affine residual correction,
\begin{equation}
    \bar v_e=\frac{v_e}{\|v_e\|_2},\qquad
    q=\bar v_e+U(D\bar v_e)+b,\qquad
    \label{eq:transform}
\end{equation}

\noindent where $D\in\mathbb{R}^{r\times d}$ and $U\in\mathbb{R}^{d\times r}$ parameterize the rank-$r$ correction, and $b\in\mathbb{R}^{d}$ is the learned bias. With $d=1024$ and $r=16$, the transform contains $33{,}793$ parameters per emotion.
$U(D\bar v_e)$ is linear: 
\(D\) is randomly initialized with orthonormal rows, while \(U\) and \(b\) are zero-initialized, making the initial residual zero and preserving the input direction.
The corrected vector $q$ is normalized and rescaled: $q/\lVert q\rVert_2$ sets the direction and $s$ the magnitude, where $s_0$ is the median $\ell_2$-norm of input contrasts and $\eta$ a learned scalar.
\begin{equation}
s=s_0(1+\beta\tanh\eta),
\qquad
v_e^\star=T_\theta(v_e)
=s\left(\frac{q}{\lVert q\rVert_2}\right)
\label{eq:transform_scale}
\end{equation}
We set $\beta=0.2$, restricting the  magnitude to within $\pm20\%$ of the reference magnitude $s_0$, i.e., $0.8s_0 < \lVert T_\theta(v_e)\rVert_2 < 1.2s_0$. This allows the transform to refine the steering direction while constraining large changes in geometry and magnitude. 
The direction $\alpha v_e^\star$ is injected at layer $l$, after which the hidden-state norm is restored to its pre-injection value.

\noindent\textit{(3) Frozen expert scoring modules.}
We use three frozen expert models to provide supervision: an emotion expert measures target-emotion strength, a speaker expert assesses identity preservation relative to the reference speech, and a transcription expert evaluates consistency with the target text. 
\subsection{Learning Objective}
\label{sec:objective}

We train the learned transform using complementary objectives for emotion control, speaker preservation, and transcript preservation, together with a regularization term that constrains changes to the original steering vector.

\noindent\textbf{Monotonic emotion supervision.}
A central goal of SteerSpeech is continuous controllability: increasing the steering strength $\alpha$ should produce a corresponding increase in target-emotion intensity. We therefore train over an ordered set of steering strengths $\alpha_0 < \alpha_1 < \cdots < \alpha_K$.

\noindent Let $\ell_j \equiv \ell(\alpha_j)$ denote the target-emotion-versus-rest log-odds of speech generated at steering strength $\alpha_j$. We encourage monotonicity by requiring the increase in emotion score between adjacent steering strengths to be proportional to the corresponding increase in $\alpha$:
\[
\mathcal{L}_{\mathrm{mono}}
=
\frac{1}{K}\sum_{j=0}^{K-1}
\left[\kappa\Delta\alpha_j-\Delta \ell_j\right]_+ ,
\]
where $\Delta\alpha_j=\alpha_{j+1}-\alpha_j$, $\Delta \ell_j=\ell_{j+1}-\operatorname{sg}(\ell_j)$, $[z]_+=\max(0,z)$, $\kappa$ controls the required increase in emotion score per unit increase in steering strength, and $\operatorname{sg}(\cdot)$ denotes stop-gradient.
This hinge loss encourages the desired ordering across steering strengths. However, monotonic ordering alone admits a trivial solution in which the emotion score increases with $\alpha$ while remaining weak at all operating points. 
We therefore include a term, where $p_e(\alpha_{\max})$ denotes the target-emotion probability at the maximum 
strength and \(\lambda\) controls how strongly the model is encouraged to express the target emotion:
\begin{equation}
\mathcal{L}_{\mathrm{emotion}}
=
\mathcal{L}_{\mathrm{mono}}
-\lambda \log p_e(\alpha_{\max})
\end{equation}

\noindent\textbf{Speaker-preservation loss.}
To measure speaker changes introduced specifically by steering, we use the corresponding unsteered generation as a reference. We penalize deviations from the unsteered output’s similarity to the reference speaker. Let $e_T$, $e_U$, and $e_R$ denote the speaker embeddings of the steered output, unsteered output, and reference speech, respectively. We define
\begin{equation}
\resizebox{0.9\columnwidth}{!}{$
s_T=\cos(e_T,e_R),\qquad
s_U=\operatorname{sg}\!\left(\cos(e_U,e_R)\right),\qquad
\mathcal{L}_{\mathrm{speaker}}=\left|s_T-s_U\right|
$}
\end{equation}

\noindent\textbf{Transcript-preservation loss.}
Given generated audio \(x\) and target transcript tokens \(y_{1:T}\),
we use the differentiable teacher-forced token negative log-likelihood,
$\mathcal{L}_{\mathrm{ASR}} = \allowbreak
-\frac{1}{T}\sum_{t=1}^{T}\allowbreak
\log P(y_t \mid y_{<t},x).
$

\noindent\textbf{Regularization loss.}
We constrain directional and magnitude drift with
$\mathcal{L}_{\mathrm{reg}} = 0.01\left[1-\cos(q,\bar v_e)\right] + 0.01\eta^2
$; computed once per unique contrast vector after averaging per-row losses.

\noindent\textbf{Combined learning objective.}
The per-sample losses are combined as follows: the resulting objective is backpropagated only to the learned steering transform.
\begin{equation}
    \label{eq:total_loss}
    \mathcal{L}_{\mathrm{update}}
    =
        \mathcal{L}_{\mathrm{emotion}}
        +
        32\mathcal{L}_{\mathrm{speaker}}
        +
        0.2\mathcal{L}_{\mathrm{ASR}}
    +
    \mathcal{L}_{\mathrm{reg}}
\end{equation}

\begin{table*}[t]
\centering
\caption{End-to-end comparison across target emotions.}
\label{tab:emotion_results}
\small
\setlength{\tabcolsep}{3pt}
\renewcommand{\arraystretch}{0.95}
\resizebox{\textwidth}{!}{%
\begin{tabular}{ll|ccccc|ccccc|ccccc}
\toprule
& & \multicolumn{5}{c|}{\textbf{Angry}}
& \multicolumn{5}{c|}{\textbf{Happy}}
& \multicolumn{5}{c}{\textbf{Sad}} \\

\textbf{Method}
& \textbf{Subset}
& Emo.$\uparrow$
& Top-1$\uparrow$
& Drift$\downarrow$
& WER$\downarrow$
& UTMOS$\uparrow$
& Emo.$\uparrow$
& Top-1$\uparrow$
& Drift$\downarrow$
& WER$\downarrow$
& UTMOS$\uparrow$
& Emo.$\uparrow$
& Top-1$\uparrow$
& Drift$\downarrow$
& WER$\downarrow$
& UTMOS$\uparrow$ \\
\midrule

Emotion reference
& Pooled
& 0.617
& 0.775
& \textbf{0.189}
& \textbf{4.9}
& \textbf{4.077}
& 0.333
& 0.394
& \textbf{0.172}
& \textbf{2.4}
& \textbf{4.131}
& 0.033
& 0.019
& 0.158
& \textbf{6.0}
& \textbf{4.256} \\

Naive
& Pooled
& 0.651
& 0.755
& 0.233
& 6.0
& 2.540
& 0.487
& 0.615
& 0.247
& 7.1
& 2.999
& 0.126
& 0.100
& \textbf{0.156}
& 8.0
& 3.452 \\
\midrule

\multirow{4}{*}{\textbf{\sys{} (ours)}}
& Seen
& 0.848
& 0.942
& 0.188
& 6.1
& 3.209
& 0.589
& 0.825
& 0.268
& 8.0
& 3.211
& 0.255
& 0.283
& 0.191
& 8.8
& 3.503 \\

& Unseen
& 0.807
& 0.925
& 0.161
& 6.1
& 3.142
& 0.554
& 0.800
& 0.275
& 15.5
& 2.955
& 0.166
& 0.175
& 0.151
& 10.6
& 3.752 \\

& Accented
& 0.824
& 0.975
& 0.252
& 2.3
& 2.980
& 0.302
& 0.425
& 0.208
& 4.7
& 4.146
& 0.246
& 0.250
& 0.192
& 4.2
& 3.513 \\
\cmidrule{2-17}

& \textbf{Pooled}
& \textbf{0.835}
& \textbf{0.945}
& 0.195
& 5.1
& 3.149
& \textbf{0.524}
& \textbf{0.740}
& 0.258
& 9.7
& 3.347
& \textbf{0.235}
& \textbf{0.255}
& 0.183
& 7.8
& 3.555 \\
\bottomrule
\end{tabular}
}
\vspace{-5mm}
\end{table*}

\vspace{-3mm}
\subsection{Methodology}
\label{sec:training}
Training $T_\theta$ is challenging because expert supervision is available only after autoregressive generation and decoding, while discrete codec-token sampling is non-differentiable; thus, gradients from the emotion, speaker, and transcription experts cannot directly propagate back to the transform. 

We address this with a two-pass pipeline: (1) we extract emotion-to-neutral activation differences from matched utterances and transform each contrast using $T_\theta$; (2) in Pass 1, for each contrast and steering strength $\alpha$, we inject the transformed direction at layer $l$ and autoregressively sample discrete codec tokens using standard inference; (3) in Pass 2, we replay the sampled sequence with teacher forcing \cite{williams1989learning} and a straight-through estimator (STE) \cite{bengio2013estimating}, retaining the discrete codec tokens in the forward pass while propagating gradients through their continuous representations, which a differentiable decoder converts to waveforms; and (4) frozen experts evaluate emotion intensity, speaker preservation, and transcript consistency across multiple $\alpha$ values. This joint evaluation encourages monotonic target-emotion increase while constraining speaker and content drift. The combined loss in Eq.~\ref{eq:total_loss} updates only $T_\theta$, while the TTS backbone, decoder, and experts remain frozen. At inference, the expert evaluators and differentiable replay path are removed.

\section{results and analysis}
\label{sec:er}
\subsection{Experimental Setup}
\noindent\textbf{Datasets.}
We use the English subset of the Emotional Speech Dataset (ESD) \cite{zhou2021esd}, containing 350 parallel utterances per speaker from 
10 native English speakers;
Audio is sampled at 16 kHz and 
average duration is 5-10 seconds. 
Evaluation covers (1) seen: unseen utterances from in-domain speakers (2) unseen: an unseen speaker split with two held-out English speakers, and (3) accented: cross-corpus generalization on four non-native Mandarin speakers from L2-ARCTIC \cite{zhao18b_interspeech}. We reserve a separate held-out speaker split for validation, distinct from the train and test splits.

\noindent\textbf{Models.} We use Qwen3-TTS-0.6B \cite{hu2026qwen3tts}, a popular TTS model with Qwen language model backbone. Our rank-16 transform has 33K parameters. The expert models used during training were emotion2vec for emotion classification~\cite{ma2024emotion2vec}, WavLM for speaker verification~\cite{chen2022wavlm}, and Whisper for ASR~\cite{radford2023whisper}. For evaluation, we used an XLSR-53-based emotion classifier~\cite{conneau2021xlsr}, ECAPA-TDNN for speaker verification~\cite{desplanques2020ecapa}, wav2vec~2.0 for ASR-based WER~\cite{baevski2020wav2vec2} and UTMOS~\cite{saeki2022utmos} for speech-quality assessment.

\noindent\textbf{Implementation Details.}
We use temperature \(0.9\), top-\(k=50\), top-\(p=1.0\), a fixed seed, and steer at layer 15. For monotonic supervision, we use 
$\alpha\in\{0.3,0.4,\ldots,1.0\}$. The transform is trained with AdamW (batch 32, learning rate \(10^{-3}\), weight decay \(10^{-4}\), and 5\% warmup with cosine decay). For evaluation, we construct one global difference of means contrast per emotion from held-out speakers and transform it using \(T_\theta\).

\noindent\textbf{Baselines.}
We compare against the following baselines:
(1) \textit{Emotion reference}: generates emotional speech using a reference emotional utterance as conditioning signal 
\cite{zhang2023iemotts}
(2) \textit{Naive difference-of-means (or Naive)}: Constructs a steering direction by subtracting the mean neutral representation from the mean representation of the target emotion. 
(3) \textit{\sys{} (ours)}: Our system, evaluated on the three splits described above: seen, unseen and accented. We report both split-level and pooled performance. Additional baselines were omitted because they require model-specific reimplementation and cross-paper results are not directly comparable.

\noindent\textbf{Evaluation Metrics.}
\textit{Objective} metrics include (1) Emotion confidence ($\uparrow$) (2) Top-1 accuracy (\(\uparrow\)), the percentage of samples classified as the target emotion (3) Speaker drift ($\downarrow$) calculated as how much the steered  speech deviates from the unsteered synthesized speech to account for any model bias and (4) Word error rate (WER) ($\downarrow$) between synthesized speech transcript and reference text and (5) UTMOS ($\uparrow$), predicted mean opinion score (MOS) assessing the naturalness of synthesized speech.
\textit{Subjective} metrics include (1) emotion preference (2) speaker-ID preference, and (3) naturalness in Survey 1; and (1) speaker-ID preservation and (2) perceived emotion increase in Survey 2. All subjective metrics are $\uparrow$.

\begin{table}[t]
    \centering
    \caption{Subjective Evaluation: A/B preference (\%). Ours show maximum emotion intensity with competitive naturalness.}
    \label{tab:subjective_results}
    \footnotesize
    \setlength{\tabcolsep}{3.5pt}
    \renewcommand{\arraystretch}{0.95}
    \resizebox{\columnwidth}{!}{%
    \begin{tabular}{@{}lccc@{}}
        \toprule
        \textbf{Comparison}
        & \makecell{\textbf{Emotion}\\\textbf{preference} $\uparrow$}
        & \makecell{\textbf{Speaker ID}\\\textbf{preference} $\uparrow$}
        & \textbf{Naturalness} $\uparrow$ \\
        \midrule

        \sys{} (ours) vs. Naive
        & 78.1/21.9
        & 41.2/58.8
        & 47.7/52.3 \\

        \sys{} (ours) vs. Emotion ref.
        & 96.8/3.2
        & 12.6/87.4
        & 16.2/83.8 \\

        Naive vs. Emotion ref.
        & 83.5/16.5
        & 20.3/79.7
        & 21.2/78.8 \\
        \bottomrule
    \end{tabular}%
    }
    \vspace{-3mm}
\end{table}

\begin{table}[t]
    \centering
    \caption{Subjective Evaluation: Effect of steering strength. \sys{} better preserves speaker identity at high intensity ($\alpha$).}
    \label{tab:steering_strength_results}
    \footnotesize
    \setlength{\tabcolsep}{2.5pt}
    \renewcommand{\arraystretch}{0.95}
    \resizebox{\columnwidth}{!}{%
    \begin{tabular}{@{}lrrr@{\hspace{6pt}}rrr@{}}
        \toprule
        & \multicolumn{3}{c}{\textbf{Speaker ID preserved (\%)}}
        & \multicolumn{3}{c}{\textbf{Anger increase (\%)}} \\
        \cmidrule(lr){2-4}
        \cmidrule(l){5-7}
        \makecell[l]{\textbf{Intensity ($\alpha$)}}
        & \textbf{Ours}
        & \textbf{Naive}
        & \textbf{$\Delta$}
        & \textbf{Ours}
        & \textbf{Naive}
        & \textbf{$\Delta$} \\
        \midrule
        $0.0 \rightarrow 0.3$
        & 89.2 & 84.2 & $+5.0$  & 58.3 & 47.5 & $+10.8$ \\
        $0.3 \rightarrow 0.5$
        & 87.5 & 84.9 & $+2.6$  & 80.8 & 67.2 & $+13.6$ \\
        $0.5 \rightarrow 0.7$
        & 65.8 & 67.5 & $-1.7$  & 71.7 & 62.5 & $+9.2$ \\
        $0.7 \rightarrow 0.9$
        & 66.7 & 46.6 & $+20.1$ & 83.3 & 41.4 & $+42.0$ \\
        $0.9 \rightarrow 1.0$
        & 69.2 & 47.4 & $+21.8$ & 58.3 & 36.2 & $+22.1$ \\
        \bottomrule
    \end{tabular}%
    }
    \vspace{-5mm}
\end{table}

\noindent\textbf{Subjective Evaluation.}
We conducted two blinded anger listening studies with 55 participants and at least 16 labels per comparison. Survey 1 used 60 A/B comparisons to assess preferences between \sys{} and baselines across all splits. Survey 2 compared five adjacent steering-strength ($\alpha$) transitions for \sys{} and Naive using identical speakers and text. Listeners judged whether higher strength increased anger while preserving speaker identity.

\vspace{-3mm}
\subsection{End-to-End Results}
\begin{figure}[t]
    \centering
    \includegraphics[width=0.85\columnwidth]{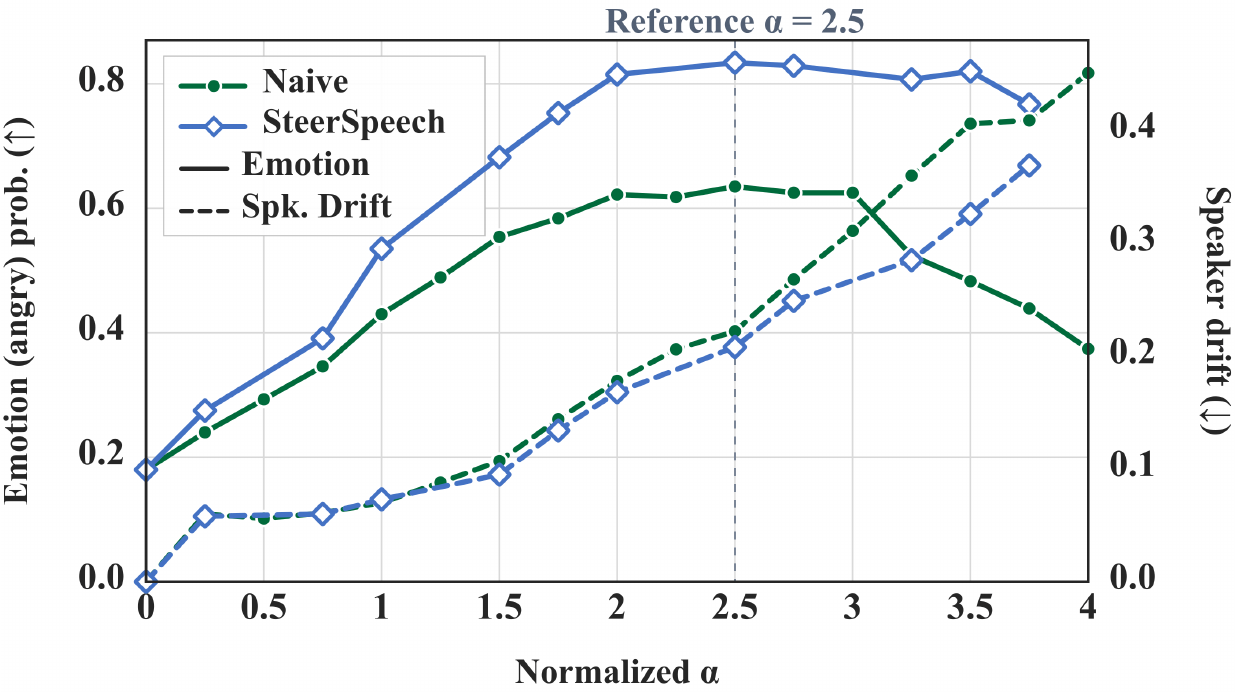}
     \vspace{-3mm}
    \caption{
    Tradeoff between target-emotion (angry) score and speaker drift at increasing steering strength ($\alpha$). 
    Ours show stronger emotion scores with lower speaker drift, especially at higher $\alpha$, whereas Naive steering shows increasing drift \& declining emotion intensity.
    }
    \label{fig:pareto_tradeoff}
    \vspace{-5mm}
\end{figure}

\noindent\textbf{Overall performance.}
Table~\ref{tab:emotion_results} compares end-to-end performance at each method's best-performing $\alpha$ from grid search (peak achievable performance); Fig.~\ref{fig:pareto_tradeoff} confirms the advantage holds at matched $\alpha$. \sys{} improves target-emotion scores by 3.7-18.4 percentage points (pp) and top-1 accuracy by 12.5-19.0pp across emotions. 
Others are emotion-dependent: anger reduces both speaker drift (by 0.038) and WER (by 0.9pp), whereas happy and sad incur modestly higher speaker drift (+0.011/+0.027), with sadness exhibiting the lowest emotion score (0.235). 
This can be explained by emotion-specific acoustic cues: anger exhibits salient high-arousal characteristics that may be easier to steer, whereas happiness relies on subtler valence cues and sadness more strongly on temporal and rhythmic patterns;
similar trend appears in recent work 
\cite{du2026sparse}. \sys{} UTMOS score improves over Naive by 0.103-0.609 across emotions. The emotion reference baseline has the weakest emotion intensity (\sys{} is 1.35$\times$-7.12$\times$ higher); its higher naturalness and lower speaker drift largely reflect smaller changes from the original speech.

Survey 1 (Table \ref{tab:subjective_results}) shows that, at $\alpha=1.0$, \sys{} is preferred for emotion intensity (78.1\% vs. 21.9\%), while Naive is preferred for speaker-ID preservation (58.8\% vs. 41.2\%), revealing an emotion–identity trade-off.
Survey 2 (Table \ref{tab:steering_strength_results}) shows that \sys{} better preserves speaker identity at higher intensities, exceeding Naive by 20.1 and 21.8pp and in perceived anger increases by 42.0 and 22.1 pp,
demonstrating higher identity preservation at high emotion intensity.

\noindent\textbf{Generalization.} \sys{} generalizes to unseen and accented speakers. For anger, unseen speakers show modest reductions in emotion and Top-1 accuracy (0.848 to 0.807; 0.942 to 0.925), lower speaker drift (0.188 to 0.161), and unchanged WER. Accented speakers achieve 0.824 emotion, 0.975 Top-1, 2.3\% WER, and 0.252 speaker drift.
For happy and sad, unseen and accented speakers retain 51–97\% of seen-speaker emotion and Top-1.
Detailed subjective results confirm this: anger is perceived monotonically in 78.5\%, 93.6\%, and 65.6\% of comparisons for seen, unseen, and accented speakers, respectively; accented speakers show the highest identity (55.6\%) and naturalness (61.7\%).

\noindent\textbf{Ablation.} We use angry as a representative emotion in ablation.

\noindent\textit{Emotion controllability/tradeoff.}
\sys{} achieves a consistently better trade-off between emotion intensity and speaker drift than the Naive baseline (Fig. \ref{fig:pareto_tradeoff}). The optimal range is approximately $\alpha=2.0$-$2.5$; at $\alpha=2.5$, \sys{} reaches a peak emotion probability of 0.834 with slightly lower drift than Naive. Beyond $\alpha\approx3$, drift increases for both methods, but Naive’s emotion intensity declines sharply while \sys{} remains comparatively robust. At matched emotion intensity, ours reduces speaker drift by 12–67\% versus Naive; WER falls 12–21\% on seen/accented speech but rises 3\% on unseen speech.

\noindent\textit{Component contribution.} On a validation subset, adding $L_{\text{speaker}}$ to $L_{\text{emotion}}$ reduced speaker drift by 0.042, while $L_{\text{ASR}}$ reduced WER by 3.82 pp. Combining both provides a balanced operating point, highlighting the complementary roles of the loss components.

\noindent\textit{Matrix vs. bias.}
We ablate the trained low-rank 
correction \(U(D\bar v_e)\) and bias \(b\) in Eq.~\ref{eq:transform} at inference, without retraining and with matched steering norms. At \(\alpha=0.7\), unseen-speaker anger accuracy is 92.5\% for the full transform, 85.0\% with only \(U(D\bar v_e)\), and 80.0\% with only \(b\); on seen speakers, either component nearly matches the full transform, indicating overlapping, split-dependent benefits.

\vspace{-2mm}
\section{Conclusion}
\label{sec:c}
In this work, we introduce \sys{}, a lightweight framework for continuous, monotonic emotion control with robust generalization to unseen and accented speakers. Unlike prior steering methods that inject a fixed or unsupervised direction, \sys{} jointly trains a low-rank transform with emotion, speaker, and transcription experts so that emotion intensity increases monotonically with steering strength while speaker and content drift stay constrained. Our results show that this supervision enables reliable inference-time emotion control without modifying the TTS backbone. Future work will extend \sys{} to additional TTS backbones and emotions.

\bibliographystyle{IEEEbib}
\bibliography{strings,refs}

\end{document}